\documentclass{PoS}
\usepackage[mathscr]{euscript}
\usepackage[all,color]{xy}
\usepackage{relsize}
\usepackage{tikz}
\usepackage{url}
\usepackage{amsmath,amsthm,bm}
\usepackage{cancel}
\usepackage{accents}
\usepackage[]{graphicx}
\usepackage{mwe} 
\usepackage{caption}
\usepackage[T1]{fontenc}
\usepackage[toc,page]{appendix}

\usepackage{empheq}
\usepackage{color}
\usepackage{mathtools}

\makeatletter

\newcommand{\dd}{\mathrm{d}}

\newcommand{\bbm}{\left(\begin{matrix}}
	\newcommand{\ebm}{\end{matrix}\right)}
\newcommand{\beq}{\begin{eqnarray}}
\newcommand{\eeq}{\end{eqnarray}}

\makeatother

 \def\one{\mbox{1 \kern-.59em {\rm l}}}

\title{Noncommutative Gauge Theories and Gravity}

\ShortTitle{Noncommutatve gravity}

\author{\speaker{G. Manolakos}\\
        Physics Department, National Technical
University, GR-15780 Athens, Greece\\
        E-mail: \email{gmanol@central.ntua.gr}}

\author{P. Manousselis\\
        Physics Department, National Technical
University, GR-15780 Athens, Greece\\
        E-mail: \email{pman@central.ntua.gr}}
        
\author{G. Zoupanos\\
        Physics Department, National Technical
University, GR-15780 Athens, Greece\\Institute of Theoretical Physics, D-69120 Heidelberg, Germany\\Max-Planck Institut f\"ur Physik, Fohringer Ring 6, D-80805 Munchen, Germany\\ Laboratoire d' Annecy de Physique Theorique, Annecy, France\\
        E-mail: \email{George.Zoupanos@cern.ch}}        

\abstract{First, we briefly review the description of gravity theories as gauge theories in three and four dimensions. Specifically, we recall the procedure in which the results of General Relativity in three and four dimensions are recovered in a gauge-theoretic approach. Also, the procedure is applied for the case of the Weyl gravity, too. Then, after reminding briefly the formulation of gauge theories on noncommutative spaces, we review our most recent works in which gravity models are constructed as gauge theories on noncommutative spaces.}

\FullConference{Corfu Summer Institute 2018 "School and Workshops on Elementary Particle Physics and Gravity"\\
		(CORFU2018)\\
		31 August - 28 September, 2018\\
		Corfu, Greece}
 
\begin{document}
\section{Introduction}
Three out of four interactions of nature are grouped together under a common description by the Standard Model in which they are described by gauge theories. However, the gravitational interaction is not part of this picture, admitting a separate, geometric formulation, that is the theory of General relativity. In order to make contact among the two different pictures, there has been an undertaking in which gravity admits a gauge-theoretic approach, besides the geometric one \cite{Utiyama:1956sy}-\cite{Witten:1988hc}.
Pioneer in this field was Utiyama, whose work was focused on describing 4-d gravity of General Relativity as a gauge theory, localizing the Lorentz symmetry, SO(1,3) \cite{Utiyama:1956sy}. However, the results were not considered to be successful, since the inclusion of the vielbein did not happen in a convincing way. A few years later, it was Kibble \cite{Kibble:1961ba} who modified the above consideration, adopting the inhomogeneous Lorentz group (Poincar\'e group), ISO(1,3), as the gauge group in which, along with the spin connection, the vierbein were also identified as gauge fields of the theory. Nevertheless, the dynamics of General Relativity remained unretrieved, since there was no action of gauge-theoretic origin of the Poincar\'e gauge group that would be identified as the Einstein-Hilbert action. Solution to this problem was given with the consideration of an SO(1,4) gauge invariant Yang-Mills action (instead of the Poincar\'e one) along with the involvement of a scalar field in the fundamental representation of the gauge group, SO(1,4) \cite{Stelle:1979aj} (see also \cite{MacDowell:1977jt,Ivanov:1980tw,Kibble:1985sn}). The gauge fixing of this scalar field led to a spontaneous symmetry breaking, recovering the Einstein-Hilbert action. Therefore, the 4-d gravitational theory of General Relativity was successfully described as a gauge theory with the presence of a scalar field.

Moreover, in the absence of cosmological constant, 3-d Einstein gravity can be also described as a gauge theory of the 3-d Poincar\'e group, ISO(1,2). In turn, the 3-d de Sitter and Anti de Sitter groups, SO(1,3) and SO(2,2), respectively are employed, in case a cosmological constant is present \cite{Witten:1988hc}. The first part of the construction, that is the calculation of the transformation of the gauge fields (dreibein and spin connection) and the curvature tensors is similar to the 4-d case. However, the dynamic part is less tedious than that of the 4-d case. The 3-d Einstein-Hilbert action is recovered after the consideration of a Chern-Simons action functional, which is, in fact, identical to the 3-d Einstein-Hilbert's action. Thus,  3-d Einstein gravity is precisely equivalent to an ISO(1,2) Chern-Simons gauge theory.

Another contribution in this aspect is related with the gauge-theoretic approach of Weyl gravity (and supergravity) as a gauge theory of the 4-d conformal group \cite{Kaku:1977pa,Fradkin:1985am}\footnote{See also \cite{vanproeyen}.}. Proceeding in the same spirit as in the previous cases, the transformations of the gauge fields and the expressions of the various curvature tensors are obtained. The action is determined to be SO(2,4) gauge invariant of Yang-Mills type, as it is expected. Then, constraints are imposed on the curvature tensors and along with gauge fixing of the fields, the final action is actually the Weyl action. Therefore, it is understood that Weyl gravity admits a gauge-theoretic interpretation of the conformal group.     

An appropriate framework for the construction of physical theories at the high-energy regime (Planck scale), in which commutativity of the coordinates cannot be naturally assumed, is that of noncommutative geometry \cite{connes} - \cite{Gavriil:2015lka}. A very improtant feature of this framework is the potential regularization of quantum field theories and the construction of finite theories. Nevertheless, building quantum field theories on noncommutative spaces is a tedious task and, moreover, problematic ultraviolet features have been encountered \cite{filk} (see also \cite{grosse-wulkenhaar} and \cite{grosse-steinacker}). Despite that, the framework of noncommutative geometry is considered to be a suitable background for accommodating particle physics models, formulated as noncommutative gauge theories \cite{connes-lott} (see also \cite{martin-bondia, dubois-madore-kerner, madorejz}).

Also, taking into account the above correspondence between gravity and (ordinary) gauge theories, the well-established formulation of noncommutative gauge theories \cite{Madore:2000en} allows one to use it as methodology for the construction of models of noncommutative gravity. Such approaches have been considered before, see for example refs. \cite{Chamseddine:2000si}-\cite{Ciric:2016isg} and, specifically, for 3-d models, employing the Chern-Simons gauge theory formulation, see \cite{Cacciatori:2002gq}-\cite{Banados:2001xw}. The authors of the above works to which we refered make use of constant noncommutativity (Moyal-Weyl) and also use the formulation of the $\star$-product and the Seiberg-Witten map \cite{Seiberg:1999vs}.

However, besides the $\star$-product formulation, noncommutative gravitational models can be constructed using the noncommutative realization of matrix geometries \cite{Banks:1996vh,Ishibashi:1996xs}. Such approaches, specifically for Yang-Mills matrix models, were proposed in the past few years, see refs. \cite{Aoki:1998vn}-\cite{Nair:2006qg}. Also, for alternative approaches on the subject see \cite{Buric:2006di,Buric:2007zx,Buric:2007hb}, but also \cite{Aschieri:2003vyAschieri:2004vhAschieri:2005wm}. In general, formulation of noncommutative gravity implies that the noncommutative deformations break the Lorentz invariance. However, there exist specific noncommutative deformations which preserve the Lorentz invariance and the corresponding background spaces are called covariant noncommutative spaces \cite{Snyder:1946qz,Yang:1947ud}. Along these lines, in ref.\cite{Heckman:2014xha}, a noncommutative deformation of a general conformal field theory defined on 4-d dS or AdS spacetime has been employed, see also \cite{Buric:2015wta}-\cite{Steinacker:2016vgf}. 

In this proceedings contribution, our recent contributions in the above field of noncommutative gravity are included. First, we briefly review our proposition for a matrix model of 3-d noncommutative gravity \cite{Chatzistavrakidis:2018vfi} (see also \cite{Manolakos:2018isw,Manolakos:2018hvn}), in which the corresponding background space is the $\mathbb{R}_\lambda^3$, introduced in ref. \cite{Hammou:2001cc} (see also ref. \cite{Vitale:2014hca} for field theories on this space), which is actually the 3-d Euclidean space foliated by multiple fuzzy spheres of different radii. As explained in ref.\cite{Kovacik:2013yca}, the above fuzzy space admits an SO(4) symmetry, which is in fact the gauge group we considered. Noncommutativity implies the enlargement of the SO(4) to the U(2)$\times$U(2) gauge group, in a fixed representation, in order that the anticommutators of the generators close. In the same spirit, the Lorentz analogue of the above construction was also explored, in which the corresponding noncommutative space is the $\mathbb{R}_\lambda^{1,2}$, that is the 3-d Minkowski spacetime foliated by fuzzy hyperboloids \cite{Jurman:2013ota}. In this case too, the initial gauge group, SO(1,3), is eventually extended to GL(2,$\mathbb{C}$) in a fixed representation, for the same reasons as in the Euclidean case. In both signatures, the action proposed is a functional of Chern-Simons type and its variation produces the equations of motion. In addition, the commutative limit is considered, retrieving the expressions of the 3-d Einstein gravity. 

Second, a 4-d gravity model as a noncommutative gauge theory is constructed \cite{Manolakos:2019fle}. Motivated by Heckman-Verlinde \cite{Heckman:2014xha} who were based on Yang's early work \cite{Yang:1947ud}, we considered a noncommutative version of the 4-d de Sitter space, which is in fact a covariant fuzzy space, preserving Lorentz invariance. Some of the generators of the algebra of the final symmetry group of this space, SO(1,5), are identified as its noncommutative coordinates. As in the previous 3-d case, the final gauge group is a minimal extension of the initial, SO(1,4), specifically the SO(1,5)$\times$U(1), for the same reasons. According to the standard procedure, the corresponding gauge fields are determined and their gauge transformations and their corresponding component curvature tensors are obtained. Eventually, an action of Yang-Mills type is employed and its initial gauge symmetry breaks imposing certain conditions (constraints) on the curvature tensors and the gauge fields. The commutative limit of the model reduces the obtained expressions of gauge transformations of the fields and the ones of the tensors to those of the conformal gravity. 

The outline of the present contribution is as follows: First we briefly recall the gauge-theoretic approaches of gravitational theories mentioned above. Then, we include the necessary information about the construction of gauge theories in the noncommutative framework. Next, we review our suggestions for noncommutative gravity models in three and four dimensions. Eventually, we write down our conclusions and comment on the results. 
\section{Gauge-theoretic approach of gravity}
In this section we briefly review the gauge-theoretic approach of various gravity theories \cite{Utiyama:1956sy}-\cite{Witten:1988hc}, which consists the basis of our works in which the whole scheme is translated to the framework of noncommutativity. 
\subsection{3-d Einstein gravity}\label{3dgravity}
Let us begin with the 3-d case, in which Einstein gravity is precisely described by a Chern-Simons gauge theory of the ISO(1,2), Poincar\'e group \cite{Witten:1988hc}. Specifically for the dynamic part, the 3-d Einstein-Hilbert action is written down as: 
\begin{equation} \label{eh3}
S_{\text{EH3}}=\frac{1}{2}\int \dd^3x \,\epsilon^{\mu\nu\rho}\epsilon_{a b c}\, e_{ \mu}{}^{a}
R_{ \nu \rho}{}^{ b c}(\omega)~,
\end{equation}
which, as it will be explained, is identical to a Chern-Simons functional of ISO(1,2).

The ISO(1,2) algebra is generated by six operators, i.e. the three local translations, $P_{ a}$ and the three Lorentz transformations, $M^{ a}=\epsilon^{ a b c}M_{ b c}$, with $ a=1,2,3$ and determine the algebra through the following commutation relations:
\begin{equation}
[M_{ a},M_{ b}]=\epsilon_{ a b c}M^{ c}~,\,\,\,\,\,\,
{[}P_{ a},M_{ b}]= \epsilon_{ a b c}P^{ c}~,\,\,\,\,\,
{[}P_{ a},P_{ b}]=0~.
\end{equation}
According to the gauging procedure, the gauge connection and gauge parameter are written down as:
\begin{equation}
A_{ \mu}=e_{ \mu}{}^{ a}P_{ a}+\omega_{ \mu}{}^{ a}M_{ a}~,\,\,\,\,\,
\epsilon=\xi^{ a}P_{ a}+\lambda^{ a}M_{ a}~,
\end{equation}
where the vielbein and the spin connection are identified as the gauge fields of the theory. The above expressions along with the transformation rule of the gauge connection:
\begin{equation}
\delta A_\mu=D_\mu \epsilon=\partial_\mu+[A_\mu,\epsilon]~,\label{poincaretransformation}
\end{equation}
lead to the calculation of the gauge transformations of the component fields:
\begin{align}
\delta e_{\mu}^{~a}&=\partial_{\mu}\xi^a+\omega_{\mu}^{~ab}\xi_b-\lambda^a_{~b}e_{\mu}^{~b}~,
\\
\delta \omega_{\mu}^{~ab}&=\partial_{\mu}\lambda^{ab}+\lambda^a_{~c}\omega_\mu^{~bc}-\lambda^b_{~c}\omega_\mu^{~ac}~
\end{align}
and the component curvature tensors:
\begin{align}
T_{\mu\nu}^{~~a}&=\partial_\mu e_\nu^{~a}-\partial_\nu e_\mu^{~a}-\omega_\mu^{~ab}e_{\nu b}+\omega_\nu^{~ab}e_{\mu b}~,\nonumber \\
R_{\mu\nu}^{~~ab}&=\partial_\mu\omega_\nu^{~ab}-\partial_\nu\omega_\mu^{~ab}-\omega_\mu^{~ac}\omega_{\nu c}^{~~b}+\omega_\nu^{~ac}\omega_{\mu c}^{~~b}~,\label{poincarecomponenttensors}
\end{align}
starting from the defining relation of the curvature two-form, $R_{\mu\nu}$:
\begin{equation}
R_{\mu\nu}=[D_\mu,D_\nu]=\partial_\mu A_\nu - \partial_\nu A_\mu+[A_{\mu},A_{\nu}]\label{poincarecurvature}
\end{equation}
and expanding it on the generators of the algebra as:
\begin{equation}
R_{\mu\nu}=T_{\mu\nu}^{~~a}P_a+\frac{1}{2} R_{\mu\nu}^{~~ab}M_{ab}~.\label{poincarecurvatureexpansion}
\end{equation}
Also, consideration of a Chern-Simons action functional leads to the Einstein-Hilbert action, \eqref{eh3}. In addition, 3-d gravitational theory with cosmological constant is also described in a gauge-theoretic approach. In this case, the gauge groups considered are the 3-d dS or AdS groups, SO(1,3) and SO(2,2), respectively, depending on the sign of the constant. The procedure of the construction of the gauge theory is the same, obtaining results that generalize the above of the ISO(1,2) case, because of the difference in the right hand side of the commutator of the translations, which is now non-zero:   
\begin{equation} 
[P_{ a},P_{ b}]=\lambda M_{ a b}~.
\end{equation}
\subsection{4-d Einstein gravity}\label{einsteingravity}
Now, gravitational interaction in four dimensions is described by General Relativity, which consists a solid and successful theory, having passed many tests since its early days. Its formulation is geometric, differentiating it from the rest interactions, which are described as gauge theories. Aiming at a connection between gravitational and the other interactions, the undertaking of the gauge-theoretic approach of gravity took place \cite{Utiyama:1956sy}-\cite{Kibble:1985sn}. Here we mention briefly the main features of this gauge-theoretic approach of the 4-d Einstein's gravity. 

First, for this gauge-theoretic approach of 4-d gravity, the vierbein formulation of General Relativity has to be considered. The gauge group is reasonably chosen to be the ISO(1,3) (Poincar\'e group), since it is the isometry group of the 4-d Minkowski spacetime. The generators of the corresponding algebra satisfy the following commutation relations, which in fact determine the algebra:
\begin{equation}
[M_{ab},M_{cd}]=4\eta_{[a[c}M_{d]b]}~,\,\,\,\,\,
[P_a,M_{bc}]=2\eta_{a[b}P_{c]}~,\,\,\,\,\,
[P_a,P_b]=0~,
\end{equation}
where $\eta_{ab}=\mathrm{diag}(-1,1,1,1)$ is the metric tensor of the 4-d Minkowski spacetime, $M_{ab}$ are the generators of the Lorentz group (Lorentz transformations) and $P_a$ are the generators of the local translations. According to the standard gauging procedure, that is the one followed in the 3-d case described in the previous section, the gauge potential, $A_\mu$, is defined and it is expressed as a decomposition on the generators of the Poincar\'e algebra, as: 
\begin{equation}
A_{\mu}(x)=e_{\mu}{}^a(x)P_a+\frac{1}{2}\omega_{\mu}{}^{ab}(x)M_{ab}~.\label{connection}
\end{equation}
The functions accompanying the generators of the algebra in the above decomposition are identified as the gauge fields of the theory and, specifically in this case, they are identified as the vierbein, $e_\mu^{~a}$ and the spin connection, $\omega_\mu^{~ab}$, which correspond to the translations, $P_a$, and Lorentz generators, $M_{ab}$, respectively. The consideration of the vierbein as a gauge field implies the mixing between the internal symmetry and spacetime making this kind of construction special, compared to the ordinary gauge theories. The gauge connection transforms according to the following rule: 
\begin{equation}
\delta A_{\mu}=\partial_{\mu}\epsilon+[A_{\mu},\epsilon]~,\label{transformconnection}
\end{equation}
where $\epsilon=\epsilon(x)$ is the gauge transformation parameter which is also expanded on the generators of the algebra:
\begin{equation}
\epsilon(x)=\xi^a(x) P_a+\frac{1}{2} \lambda^{ab}(x)M_{ab}~.\label{parameter}
\end{equation}
Combination of the equations \eqref{connection} and \eqref{parameter} with \eqref{transformconnection} lead to the expressions of the transformations of the gauge fields: 
\begin{align}
\delta e_{\mu}{}^a&=\partial_{\mu}\xi^a+\omega_{\mu}{}^{ab}\xi_b-\lambda^a{}_be_{\mu}{}^b~,
\\
\delta \omega_{\mu}{}^{ab}&=\partial_{\mu}\lambda^{ab}-2\lambda^{[a}{}_c\omega_{\mu}{}^{cb]}~.
\end{align}
The corresponding field strength tensor of the gauge theory is defined as:
\begin{equation}
R_{\mu\nu}(A)=2\partial_{[\mu}A_{\nu]}+[A_{\mu},A_{\nu}]~\label{usualformula}
\end{equation}
and is expanded on the generators as it is valued in the algebra:
\begin{equation}
R_{\mu\nu}(A)=R_{\mu\nu}{}^a(e)P_a+\frac{1}{2} R_{\mu\nu}{}^{ab}(\omega)M_{ab}~,\label{expansion_of_curvature}
\end{equation}
where $R_{\mu\nu}{}^a(e)$ and $R_{\mu\nu}{}^{ab}(\omega)$ are the curvatures associated to the component gauge fields, identified as the torsion and curvature, respectively. Replacement of the equations \eqref{connection} and \eqref{expansion_of_curvature} in the \eqref{usualformula} leads to their explicit expressions: 
\begin{align}
R_{\mu\nu}{}^a(e)&=2\partial_{[\mu}e_{\nu]}{}^a-2\omega_{[\mu}{}^{ab}e_{\nu]b}~,\\
R_{\mu\nu}{}^{ab}(\omega)&=2\partial_{[\mu}\omega_{\nu]}{}^{ab}-2\omega_{[\mu}{}^{ac}\omega_{\nu]c}{}^b~.\label{curvtwoform}
\end{align}
Moving on with the dynamic part of the theory, the most reasonable choice is an action of Yang-Mills type, being invariant under the ISO(1,3) gauge group. However, the aim is to result with the Einstein-Hilbert action, which is Lorentz invariant and, therefore, the Poincar\'e symmetry of the initial action has to be broken to the Lorentz. This can be carried out through a spontaneous symmetry breaking, induced by a scalar field which belongs to the fundamental representation of the SO(1,4) \cite{Stelle:1979aj, Ivanov:1980tw}, that is also included in the theory. The choice of the 4-d de Sitter group is an alternative and preferred choice to that of the Poincar\'e group, since all generators of the algebra can be considered on equal footing. The spontaneous symmetry breaking leads to the breaking of the translational generators, resulting to a theory with vanishing torsion constraint and a Lorentz invariant action involving the Ricci scalar (and a topological Gauss-Bonnet term), that is Einstein-Hilbert action. Concluding, Einstein's 4-d gravity theory is not equivalent to a pure Poincar\'e gauge theory but to an SO(1,4) gauge theory with the inclusion of a scalar field and the addition of an appropriate potential term in the Lagrangian, which leads to a spontaneous symmetry breaking. 

An alternative way to obtain an action with Lorentz symmetry, is to impose that the action is invariant only under the Lorentz symmetry and not under the total Poincar\'e symmetry with which one begins. This means that the curvature tensor related to the translations has to be zero, that is the imposition of the torsionless condition, that is a constraint that is necessary for resulting with an action with Lorentz symmetry. Solution of this constraint leads to a relation of the spin connection with the vielbein: 
\begin{equation}
\omega_\mu^{~ab}=\frac{1}{2}e^{\nu a}(\partial_\mu e_\nu^{~b}-\partial_\nu e_\mu^{~b})-\frac{1}{2}e^{\nu b}(\partial_\mu e_\nu^{~a}-\partial_\nu e_\mu^{~a})-\frac{1}{2}e^{\rho a}e^{\sigma b}(\partial_\rho e_{\sigma c}-\partial_\sigma e_{\rho c})e_\mu^{~c}\,. \label{spin-viel}
\end{equation} 
However, straightforward consideration of an action of Yang-Mills type with Lorentz symmetry, would lead to an action involving the $R(M)^2$ term, which is not the correct one, since the target is the Einstein-Hilbert action. Also, such an action would imply the wrong dimensionality (zero) of the coupling constant of gravity. In order to result with the Einstein-Hilbert action, which includes a dimensionful coupling constant, the action has to be considered in an alternative, non-straightforward way, that is the construction of Lorentz invariants out of the quantities (curvature tensor) of the theory. The one that is built by certain contractions of the curvature tensor is the correct one, ensuring the correct dimensionality of the coupling constant, and is identified as the Ricci scalar and the corresponding action is eventually the Einstein-Hilbert action.
 
\subsection{4-d Conformal gravity leading to Weyl or Einstein gravity}\label{weylgravity}
Besides the above, also Weyl gravity admits a gauge-theoretic formulation, specifically of the 4-d conformal group, SO(2,4). In this case, too, the transformations of the fields and the expressions of the curvature tensors are obtained in a straightforward way. The action that is considered initially is an SO(2,4) invariant action of Yang-Mills type which breaks by the imposition of constraints on the curvature tensors. After the consideration of the constraints, the resulting action of the theory is identical to the scale invariant Weyl action \cite{Kaku:1977pa,Fradkin:1985am,vanproeyen} (see also \cite{cham-thesis, Chamseddine:1976bf}). 

Let us start with the identification of the generators of the conformal algebra of SO(2,4). The fifteen generators are the local translations, ($P_a$), the Lorentz transformations, ($M_{ab}$), the conformal boosts, ($K_a$) and the dilatations, ($D$). The algebra of SO(2,4) is determined by the commutation relations of the above generators:
\begin{align}
[M_{ab},M^{cd}]&=4M_{[a}^{~[d}\delta_{b]}^{c]}\,,\quad [M_{ab},P_c]=2P_{[a}\delta_{b]c}\,,\quad [M_{ab},K_c]=2K_{[a}\delta_{b]c}\,,\nonumber \\
[P_a,D]&=P_a\,,\quad [K_a,D]=-K_a\,,\quad [P_a,K_b]=2(\delta_{ab}D-M_{ab})\,,
\end{align}  
where $a,b,c,d=1\ldots 4$. According to the gauging procedure, the gauge potential of the theory is defined and is expanded on the various generators as:
\begin{equation}
A_{\mu} = e_{\mu}^{~a}P_{a} + \frac{1}{2}\omega_{\mu}^{~ab}M_{ab} + b_{\mu}D + f_{\mu}^{~a}K_{a}\,,\label{connection-conformal}
\end{equation}
in which a gauge field has been associated with each generator. In this case, too, the vierbein and the spin connection are identified as gauge fields of the theory. The transformation rule of the gauge connection, \eqref{connection-conformal}, is:
\begin{equation}
\delta_{\epsilon}A_{\mu} = D_{\mu}\epsilon = \partial_{\mu}\epsilon+ [A_{\mu},\epsilon]\,,\label{commrule}
\end{equation}
where $\epsilon$ is a gauge transformation parameter valued in the Lie algebra of the SO(2,4) group and for this reason it can be written as:
\begin{equation}
\epsilon = \epsilon_{P}^{~a} P_{a} + \frac{1}{2}\epsilon_{M}^{~~ab} M_{ab} + \epsilon_{D}D+ \epsilon_{K}^{~a}K_{a}\,.\label{parameter-conformal}
\end{equation}
Combination of the equations \eqref{commrule}, \eqref{connection-conformal} and \eqref{parameter-conformal}, leads to the expressions of the transformations of the gauge fields of the theory:
\begin{align}
\delta e_{\mu}^{~a} &= \partial_\mu\epsilon_P^{~a}+2ie_{\mu b}\epsilon_M^{~ab}-i\omega_\mu^{~ab}\epsilon_{Pb}-b_\mu\epsilon_K^{~a}+f_\mu^{~a}\epsilon_D\,, \nonumber \\
\delta \omega_{\mu}^{~ab} &= \frac{1}{2}\partial_\mu\epsilon_M^{~ab}+4ie_\mu^{~a}\epsilon_P^{~b}+\frac{i}{4}\omega_\mu^{~ac}\epsilon_{M~c}^{~~b}+if_\mu^{~a}\epsilon_K^{~b}\,, \nonumber \\
\delta b_{\mu} &= \partial_\mu\epsilon_D-e_\mu^{~a}\epsilon_{Ka}+f_\mu^{~a}\epsilon_{Pa}\,, \nonumber \\
\delta f_{\mu}^{~a} &= \partial_\mu\epsilon_K^{~a}+4ie_\mu^{~a}\epsilon_D-i\omega_\mu^{~ab}\epsilon_{Kb}-4ib_\mu\epsilon_P^{~a}+if_\mu^{~b}\epsilon_{M~b}^{~~a}\,. \label{conformaltrans}
\end{align}
The field strength tensor is defined by the following relation:
\begin{equation}
R_{\mu\nu} = 2 \partial_{[\mu} A_{\nu]} - i [ A_{\mu} ,A_{\nu} ]\label{fieldstrengthconformal}
\end{equation}
and is expanded on the generators as:
\begin{equation}
R_{\mu\nu}=\tilde{R}_{\mu\nu}^{~~a}P_a+\frac{1}{2}R_{\mu\nu}^{~~ab}M_{ab}+R_{\mu\nu}+R_{\mu\nu}^{~~a}K_a~.\label{conformalexpansion}
\end{equation}
Combining the equation \eqref{fieldstrengthconformal} and \eqref{conformalexpansion}, the expressions of the component curvature tensors are obtained:
\begin{align}
R_{\mu\nu}^{~~~a}(P) &= 2 \partial_{[\mu}e_{\nu]}^{~~a} + f_{[\mu}^{~~a}b_{\nu]} +  e^{~~b}_{[\mu} \omega_{\nu]}^{~~ac} \delta_{bc}, \nonumber \\
R_{\mu\nu}^{~~~ab}(M) &= \partial_{[\mu} \omega_{\nu]}^{~~ab} + \omega_{[\mu}^{~~ca} \omega_{\nu]}^{~~db} \delta_{cd} + e_{[\mu}^{~~a}e_{\nu]}^{~~b} + f_{[\mu}^{~~a}f_{\nu]}^{~~b}, \nonumber \\
R_{\mu\nu}(D) &= 2 \partial_{[\mu}b_{\nu]} + f_{[\mu}^{~~a}e_{\nu]}^{~~b}\delta_{ab}, \nonumber \\
R_{\mu\nu}^{~~~a}(K) &= 2 \partial_{[\mu}f_{\nu]}^{~~a} + e_{[\mu}^{~~a}b_{\nu]} + f_{[\mu}^{~~b}\omega_{\nu]}^{~~ac}\delta_{bc}\,. \label{conformalcurvatures}
\end{align}
Regarding the action, at first it is taken to be an SO(2,4) invariant of Yang-Mills type. The initial, SO(2,4), symmetry gets broken by the imposition of certain constraints, \cite{Kaku:1977pa,Fradkin:1985am, vanproeyen}, that is the torsionless condition, $R(P)=0$ and an additional constraint on $R(M)$. The two constraints admit an algebraic solution leading to expressions of the fields $\omega_\mu^{~ab}$ and $f_\mu^{~a}$ in terms of the independent fields $e_\mu^{~a}$ and $b_\mu$. Also, the gauge fixing $b_\mu=0$ can be employed and, inclusion of all constraints in the initial action lead to the well-known Weyl action. 

Besides the above breaking of the conformal symmetry which led to the Weyl action, it is possible to employ an alternative breaking route, this time leading to an action with Lorentz symmetry, explicitly the Einstein-Hilbert action \cite{Chamseddine:2002fd}. From our prespective, the latter can be achieved through an alternative symmetry breaking mechanism, specifically with the inclusion of two scalar fields in the fundamental representation of the conformal group \cite{Li:1973mq}, being a generalization of the case of the breaking of the 4-d de Sitter group down to the Lorentz group by the inclusion of a scalar in the fundamental representation of SO(1,4), as explained in section \ref{einsteingravity}. The inclusion of two scalars could trigger a spontaneous symmetry breaking in a theory with matter fields and the resulting action would be the Einstein-Hilbert one, respecting Lorentz symmetry. Calculations and details on this issue will be included in a future work. 

Moreover, the argument used in the previous section in the four-dimensional Poincar\'e gravity case, that is an alternative way to break the initial symmetry to the Lorentz, can be generalized for this case of conformal gravity. Since it is desired to result with the Lorentz symmetry out of the initial SO(2,4), the vacuum of the theory is considered to be directly SO(4) invariant, which means that every other tensor, except for the $R(M)$, has to be vanishing. Setting these tensors to zero will produce the constraints of the theory leading to expressions that relate the gauge fields. In particular, in \cite{Chamseddine:2002fd}, it is argued that if both tensors $R(P)$ and $R(K)$ are simultaneously set to zero, then from the constraints of the theory it is understood that the corresponding gauge fields, $f_\mu^{~a}, e_\mu^{~a}$ are equal - up to a rescaling factor - and $b_\mu=0$.

\section{Noncommutative gauge theories}
Let us now briefy recall the basic concepts of the formulation of gauge theories on noncommutative spaces, in order to use them later for the construction of the noncommutative gravity models. For convenience, the following methodology is performed on the most typical noncommutative (fuzzy) space, the fuzzy sphere \cite{Madore:1991bw}. Obviously, the results can be easily generalized in the cases of other noncommutative spaces, too. 

Let a field $\phi(X_a)$ of the fuzzy sphere, written in terms of powers of $X_a$ \cite{Madore:2000en}, and a gauge group, G. An infinitesimal gauge transformation of $\phi(X_a)$ is: 
\begin{equation}
  \delta\phi(X)=\lambda(X)\phi(X)\,,\label{gaugetransffuzzy}
\end{equation}
where $\lambda(X)$ is the gauge transformation parameter. If $\lambda(X)$ is a function of the coordinates, $X_a$, then it is an infinitesimal Abelian transformation and G=U(1), while if $\lambda(X)$ is a P$\times$P Hermitian matrix, then the transformation is non-Abelian and the gauge group is G=U(P). The coordinates are invariant under an infinitesimal transformation of the the gauge group, G, namely $\delta X_a=0$. Now, the gauge transformation of the product of the field and a coordinate is not covariant:
\begin{equation}
  \delta(X_a\phi)=X_a\lambda(X)\phi\,,
\end{equation}
since, in general, it holds:
\begin{equation}
  X_a\lambda(X)\phi\neq\lambda(X)X_a\phi\,.
\end{equation}
Drawing lessons by the construction of ordinary gauge theories, in which a covariant derivative is defined, in the noncommutative case, the covariant coordinate, $\phi_a$, is introduced by its transformation property:
\begin{equation}
  \delta(\phi_a\phi)=\lambda\phi_a\phi\,,
\end{equation}
which is satisfied when:
\begin{equation}
  \delta(\phi_a)=[\lambda,\phi_a]\,.\label{transformationofphia}
\end{equation}
Therefore, the covariant coordinate is defined as:
\begin{equation}
  \phi_a\equiv X_a+A_a\,,\label{covariantfield}
\end{equation}
where $A_a$ is identified as the gauge connection of the noncommutative gauge theory. Combining equations \eqref{transformationofphia}, \eqref{covariantfield}, the gauge transformation of the connection, $A_a$, is obtained:
\begin{equation}
  \delta A_a=-[X_a,\lambda]+[\lambda,A_a]\,.
\end{equation}
In the above relation, the role of $A_a$ as a gauge field is demonstrated. Next, the field strength tensor, $F_{ab}$, is defined on the fuzzy sphere as:
\begin{equation}
  F_{ab}\equiv[X_a,A_b]-[X_b,A_a]+[A_a,A_b]-C^c_{ab}A_c=[\phi_a,\phi_b]-C^c_{ab}\phi_c\,,
\end{equation}
which is covariant under a gauge transformation:
\begin{equation}
  \delta F_{ab}=[\lambda,F_{ab}]\,.
\end{equation}
In the next sections, the above methodology is applied on the construction of noncommutative gravity models as gauge theories. 
\section{A 3-d noncommutative gravity model}\label{noncommutativegravitythreedims}
The solid framework for constructing noncommutative gauge theories, as described in the previous section, combined with the description of 3-d gravity as a gauge theory, section \ref{3dgravity}, gives rise to the construction of a 3-d model of noncommutative gravity. First, one has to identify an appropriate noncommutative space, on which the noncommutative gauge theory is constructed on. A suitable 3-d fuzzy space is constructed by the foliation of the 3-d Euclidean space by multiple fuzzy spheres of different radii, called $\mathbb{R}_\lambda^3$, first considered in ref.\cite{Hammou:2001cc} (see also \cite{Vitale:2014hca}). The coordinates of $\mathbb{R}_\lambda^3$, satisfy the commutation relation of the SU(2) algebra, just like the coordinates of a fuzzy sphere do. However, unlike the case of the fuzzy sphere, the generators of SU(2), to which coordinate operators are related, are not accommodated in an irreducible (higher-dimensional) representation, but in a reducible one. The employment of a  reducible representation of SU(2) means that the coordinates can be expressed as matrices in a block-diagonal form, with each block being some irreducible representation, corresponded to a fuzzy sphere of certain radius. Therefore, the Hilbert space would be:
\begin{equation}
{\mathcal{H}}=\oplus [\ell],\quad \ell=0,1/2,1,\ldots~.
\end{equation}
The three coordinates of $\mathbb{R}_\lambda^3$, $X_i$, are the operators which satisfy the following commutation relation:
\begin{equation}
[X_i,X_j]=i\lambda \epsilon_{ijk}X_k~,
\end{equation}
and are described by matrices in reducible representations of the algebra of SU(2) (cf. \cite{Hammou:2001cc}). Therefore, the coordinates, $X_i$, are allowed to be set in a reducible representation and this is equivalent to a sum of fuzzy 2-spheres of different radii. Thus, the noncommutative space can be viewed as a discrete foliation of 3-d Euclidean space by fuzzy 2-spheres, each fuzzy sphere being a leaf of the foliation\footnote{In the Lorentzian signature, an analogous construction is encountered, specifically the foliation of the 3-d Minkowski spacetime by fuzzy hyperboloids of different radii \cite{Jurman:2013ota}.} (cf. \cite{DeBellis:2010sy}).

The gauge group that is adopted for the construction of the theory is the SO(4), that is the parametrization of the symmetry of the $\mathbb{R}_\lambda^3$, as it is explained in ref.\cite{Kovacik:2013yca}. During the gauging procedure, the typical problem of the non-closure of the anticommutators of the generators of the algebra is encountered. The indicated solution for this problem is to pick a specific representation of the group to accommodate the generators and enlarge the algebra to the minimal extend, including the operators that are produced by the anticommutators. Accordingly, in the specific SO(4) case of interest, the final gauge group is the U(2)$\times$U(2) in a fixed representation, due to the inclusion of two more operators $\one$ and $\gamma_5$ to the SO(4) set of generators.   

After the determination of the fuzzy space and the gauge group, the procedure that is followed is a modification of the one of the continuous case, adjusted in the framework of noncommutative geometry. At first, the commutation and anticommutation relations of the generators of the gauge group are obtained:
\begin{eqnarray}
&& [P_a,P_b]=i\epsilon_{abc}M_c~, \quad [P_a,M_b]=i\epsilon_{abc}P_c~, \quad [M_a,M_b]=i\epsilon_{abc}M_c~,
\\
&& \{P_a,P_b\}=\tfrac{1}{2} \delta_{ab}\one~,\quad \{P_a,M_b\}=\tfrac{1}{2} \delta_{ab}\gamma_5~, \quad \{M_a,M_b\}=\tfrac{1}{2} \delta_{ab}\one~.
\end{eqnarray}
Next, in order to write down the expression of the gauge connection, a gauge field has to be introduced for each generator. Therefore, the covariant coordinate is defined as: 
\begin{equation}
{\cal X}_{\mu}= X_{\mu}\otimes i\one +e_{\mu}{}â\otimes P_a+\omega_{\mu}{}â\otimes M_a+A_{\mu}\otimes i\one+{\widetilde{A}}_{\mu}\otimes \gamma_5~~.
\end{equation}
Also, a gauge transformation parameter is introduced and it is expanded on the generators of the algebra as:
\begin{equation}
\epsilon=\xi^a\otimes P_a+\lambda^a\otimes M_a+\epsilon_0\otimes i\one+\widetilde{\epsilon}_0\otimes\gamma_5~.
\end{equation}
The above relations combined with the transformation rule of the covariant coordinate, produce the following transformations of the gauge fields:
\begin{align}
\delta e_\mu^{~a}&=-i[X_\mu+A_\mu,\xi^a]+2\{\omega_{\mu b},\xi_c\}\epsilon^{abc}+2\{e_{\mu b},\lambda^c\}\epsilon^{abc}+2i[\lambda_a,\tilde{A}_\mu]+2i[\tilde{\epsilon}_0,\omega_{\mu a}]+i[\epsilon_0,e_{\mu a}]~,\nonumber\\
\delta\omega_\mu^{~a}&=-i[X_\mu+A_\mu,\lambda^a]+2\{\omega_{\mu b},\lambda_c\}\epsilon^{abc}-\frac{1}{2}\{e_{\mu b},\xi_c\}\epsilon^{abc}+\frac{i}{2}[\xi^a,\tilde{A}_\mu]+i[\epsilon_0,\omega_\mu^{~a}]+\frac{i}{2}[\tilde{\epsilon}_0,e_\mu^{~a}]~,\nonumber\\
\delta A_\mu&=-i[X_\mu+A_\mu,\epsilon_0]-i[\xi^a,e_{\mu a}]+4i[\lambda^a,\omega_{\mu a}]-i[\tilde{\epsilon}_0,\tilde{A}_\mu]~,\nonumber\\
\delta\tilde{A}_\mu&=-i[X_\mu+A_\mu,\tilde{\epsilon}_0]+2i[\xi^a,\omega_{\mu a}]+2i[\lambda^a,e_{\mu a}]+i[\epsilon_0,\tilde{A}_\mu]~.\label{transformationsofthefieldsnoncomm3-dimgravity}
\end{align}
Also, making use of the definition of the covariant coordinate in the following defining relation of the field strength tensor:
\begin{equation}
\mathcal{R}_{\mu\nu}=[\mathcal{X}_\mu,\mathcal{X}_\nu]-i\lambda\epsilon_{\mu\nu}^{~~~\rho}\mathcal{X}_\rho\,,
\end{equation}
the corresponding component curvature tensors are obtained:
\begin{align}
T_{\mu\nu}^{~~a}&=i[X_\mu+A_\mu,e_\nu^{~a}]-i[X_\nu+A_\nu,e_\mu^{~a}]-2\epsilon^{abc}\left(\{e_{\mu b},\omega_{\nu c}\}+\{\omega_{\mu b},e_{\nu c}\}\right)\nonumber\\
&~~~+2i\left([\omega_\mu^{~a},\tilde{A}_\nu]-[\omega_\nu^{~a},\tilde{A}_\mu]\right)-i\lambda \epsilon_{\mu\nu}^{~~\rho}e_\rho^{~a}~,\\
R_{\mu\nu}^{~~a}&=i[X_\mu+A_\mu,\omega_\nu^{~a}]-i[X_\nu+A_\nu,\omega_\mu^{~a}]+\epsilon^{abc}\left(\tfrac{1}{2}\{e_{\mu b},e_{\nu c}\}-2\{\omega_{\mu b},\omega_{\nu c}\}\right)\nonumber\\
&~~~+\tfrac{i}{2}\left([e_\mu^{~a},\tilde{A}_\nu]-[e_\nu^{~a},\tilde{A}_\mu]\right)-i\lambda \epsilon_{\mu\nu}^{~~\rho}\omega_\rho^{~a}~,\\
F_{\mu\nu}&=i[X_\mu+A_\mu,X_\nu+A_\nu]-i[e_{\mu}^{~a},e_{\nu a}]+4i[\omega_\mu^{~a},\omega_{\nu a}]-i[\tilde{A}_\mu,\tilde{A}_\nu]-i\lambda \epsilon_{\mu\nu}^{~~\rho}(X_\rho+A_\rho)~,\\
\tilde{F}_{\mu\nu}&=i[X_\mu+A_\mu,\tilde{A}_\nu]-i[X_\nu+A_\nu,\tilde{A}_\mu]+2i\left([e_\mu^{~a},\omega_{\nu a}]+[\omega_\mu^{~a},e_{\nu a}]\right)-i\lambda \epsilon_{\mu\nu}^{~~\rho}\tilde{A}_\rho~.\label{componentcurvturetensorsrlammbdatothe1commatwo} 
\end{align}
Finally, the action that is proposed is of Chern-Simons type, specifically:
\begin{equation}
S=\text{Tr}i\epsilon^{\mu\nu\rho}\mathcal{X}_\mu\mathcal{R}_{\nu\rho}\,.
\end{equation}
The equations of motion are obtained after variation of the above action with respect to the various gauge fields:
\begin{equation}
T_{\mu\nu}^{~~a}=0~,\quad R_{\mu\nu}^{~~a}=0~,\quad F_{\mu\nu}=0~\quad\tilde{F}_{\mu\nu}=0~.
\end{equation}
It is worth-noting that the results of the above construction of noncommutative 3-d gravity  reduce to the ones of the continuous case in section \ref{3dgravity}.
\section{A 4-d noncommutative gravity model}
In this section, the construction of a 4-d gravity model as a noncommutative gauge theory is reviewed. First, the construction of an appropriate 4-d fuzzy space, on which the gravity model is constructed, is presented and then the features of the gravity model on this 4-d fuzzy space are explored \cite{Manolakos:2019fle}.

\subsection{Fuzzy de Sitter space}\label{fuzzydesitter}
The 4-d background space that is employed in this case is the fuzzy 4-d de Sitter space,  $\mathrm{dS_4}$. The continuous $\mathrm{dS_4}$ is defined as a submanifold of the 5-d Minkowski spacetime and can be viewed as the Lorentzian analogue of the definition of the four-sphere as an embedding in the 5-d Euclidean space. Specifically, the defining, embedding equation of $\mathrm{dS_4}$ is:
\begin{equation}
\eta^{MN}x_Mx_N=R^2\,,
\end{equation}
where $M,N=0,\ldots, 4$ and $\eta^{MN}$ is the metric tensor of the 5-d Minkowski spacetime, $\eta^{MN}=\mathrm{diag}(-1,+1,+1,+1,+1)$. In order to obtain the fuzzy analogue of this space, one has to consider its coordinates, $X_m$, to be operators that do not commute with each other:
\begin{equation}
[X_m,X_n]=i\theta_{mn}\,,\label{noncommgeneral}
\end{equation}
where the spacetime indices are $m,n=1,\ldots,4$. Analogy to the fuzzy sphere case, in which the corresponding coordinates are identified as the three (rescaled) generators of SU(2) in an (large) N-dimensional representation, implies that the right hand side, \eqref{noncommgeneral}, should be identified as a generator of the underlying algebra, ensuring covariance, that is $\theta_{mn}=C_{mn}^{~~~r}X_r$, where $C_{mnr}$ is a rescaled Levi-Civita symbol. However, in this fuzzy de Sitter case, such an identification cannot be achieved, in the sense that such an identification of the coordinate operators with generators of SO(1,4) would break Lorentz invariance, since the algebra would not be closing, i.e. $\theta_{mn}$ cannot be identified as generators into the algebra \cite{Heckman:2014xha}\footnote{For more details on this issue, see \cite{Sperling:2017dts,Kimura:2002nq}, where the same problem emerges in the construction of the fuzzy four-sphere.}. However, preservation of covariance is necessary for our purpose, therefore a group with larger symmetry, in which all operators identified as coordinates but also the noncommutativity tensor can be included in it, is considered. The enlargement of the symmetry leads to the consideration of the SO(1,5) group. Therefore, a fuzzy  $\mathrm{dS}_4$ space, with its coordinates being operators represented by N-dimensional matrices, respecting covariance, too, is obtained after the enlargement of the symmetry to the SO(1,5) \cite{Manolakos:2019fle}. In order to facilitate the construction we make use of the Euclidean signature, therefore, instead of the SO(1,5), the resulting symmetry group is considered to be that of SO(6). 
  
For the explicit formulation of the above 4-d fuzzy space, let us consider the SO(6) generators, denoted as $\mathrm{J}_{AB} = - \mathrm{J}_{BA}$, with $A,B = 1,\ldots, 6$, which satisfy the following commutation relation:
\begin{equation}
[J_{AB}, J_{CD} ] = i(\delta_{AC}J_{BD} + \delta_{BD}J_{AC} - \delta_{BC}J_{AD} - \delta_{AD}J_{BC} )\,.
\end{equation}
The above generators can be written as a decomposition in an SO(4) notation, with the component generators being identified as various operators, including the coordinates: 
\begin{equation}
J_{mn} = \tfrac{1}{\hbar} \Theta_{mn}, \ \ J_{m5} = \tfrac{1}{\lambda} X_{m}, \ \  J_{m6} = \tfrac{\lambda}{2 \hbar}P_{m} , \ \  J_{56} = \tfrac{1}{2} \mathrm{h}\,,
\end{equation}
with $m,n = 1,\ldots,4$. For dimensional reasons, an elementary length, $\lambda$, has been introduced in the above identifications, in which the coordinates, momenta and noncommutativity tensor are denoted as $X_{m}$, $P_{m}$ and $\Theta_{mn}$, respectively. The coordinate and momentum operators satisfy the following commutation relations:
\begin{equation}
[ X_{m} , X_{n} ] = i \frac{\lambda^{2}}{\hbar} \Theta_{mn}, \ \ \ [P_{m}, P_{n} ] = 4i \frac{\hbar}{\lambda^{2}} \Theta_{mn}
\end{equation}
\begin{equation}
[ X_{m}, P_{n} ]  = i \hbar \delta_{mn}\mathrm{h}, \ \ \ [X_{m}, \mathrm{h} ] = i \frac{\lambda^{2}}{\hbar} P_{m}
\end{equation}
\begin{equation}
[P_{m}, \mathrm{h} ] =4i \frac{\hbar}{\lambda^{2}} X_{m}\,.
\end{equation}
The algebra of spacetime transformations is:
\begin{equation}
[X_{m}, \Theta_{np} ] = i \hbar ( \delta_{mp} X_{n} - \delta_{mn} X_{p} )
\end{equation}
\begin{equation}
[P_{m}, \Theta_{np} ] = i \hbar ( \delta_{mp} P_{n} - \delta_{mn} P_{p} )
\end{equation}
\begin{equation}
[\Theta_{mn}, \Theta_{pq} ] = i \hbar ( \delta_{mp} \Theta_{nq} + \delta_{nq} \Theta_{mp} - \delta_{np} \Theta_{mq} - \delta_{mq} \Theta_{np} )
\end{equation}
\begin{equation}
[\mathrm{h}, \Theta_{mn} ] = 0~.
\end{equation}
In contrast to the Heisenberg algebra (see \cite{Singh:2018qzk}), the above algebra admits finite-dimensional matrices to represent the operators $X_{m}$, $P_{m}$ and $\Theta_{mn}$, therefore the kind of spacetime obtained above is a finite quantum system. Spaces like the above fuzzy $\mathrm{dS}_4$ fall into the general class of fuzzy spaces, that is the fuzzy covariant spaces \cite{Heckman:2014xha,Buric:2015wta,Barut}. 

\subsection{A noncommutative gauge theory of 4-d gravity}

In this section we review the construction of a noncommutative 4-d gravity model as a gauge theory on the fuzzy $\mathrm{dS}_4$ space of the previous section, \ref{fuzzydesitter}\footnote{For a string theory approach on such a model, see \cite{AlvarezGaume:2006bn}.}. In analogy to the 3-d translation of the gauge-theoretic description of gravity on the fuzzy space $\mathbb{R}_\lambda^3$, in section \ref{noncommutativegravitythreedims}, the same pattern is followed, this time translating the 4-d case presented in section \ref{weylgravity}.
\subsection{Determination of the gauge group and representation by $4 \times 4$ matrices}
In the previous section, the fuzzy $\mathrm{dS}_4$ space was constructed and its symmetry group was found to be the SO(6). Recalling the case of the construction of Einstein gravity as gauge theory in section \ref{einsteingravity}, in which the isometry group (the Poincar\'e group) was chosen to be the gauged, in this case the role of the gauge group will be given to the isometry group of the fuzzy $\mathrm{dS}_4$ space, namely the SO(5), viewed as a subgroup of the SO(6) group.  

However, the same problem related to the anticommutators of the generators of the algebra that emerged in the 3-d case of noncommutative gravity, section \ref{noncommutativegravitythreedims}, is encountered in this case, too \cite{Chatzistavrakidis:2018vfi,Manolakos:2018isw,Manolakos:2018hvn} (see also \cite{Aschieri1}). Specifically, the anticommutation relations of the generators of the gauge group, SO(5), produce operators that, in general, do not belong to the algebra. Therefore, to troubleshoot this problem, the representation of the generators has to be fixed and all operators produced by the anticommutators of the (fixed) generators have to be included into the algebra, identifying them as generators, too. Thus, the initial gauge group, SO(5) is extended to the SO(6)$\times$U(1) group with the generators being represented by 4$\times$4 matrices (in the 4 representation of SO(6)).

In order to obtain the specific expressions of the matrices representing the generators, the four Euclidean $\Gamma$-matrices are employed, satisfying the following anticommutation relation:
\begin{equation}
\{\Gamma_{a}, \Gamma_{b} \} = 2 \delta_{ab} \one\,,
\end{equation} 
where $a,b = 1,\ldots 4$. Also the $\Gamma_{5}$ matrix is defined as $ \Gamma_{5} = \Gamma_{1} \Gamma_{2} \Gamma_{3} \Gamma_{4} $. Therefore, the generators of the SO(6)$\times$U(1) gauge group are identified as:\\\\
a) Six generators of the Lorentz transformations:
$ \mathrm{M}_{ab} = - \tfrac{i}{4} [\Gamma_{a} , \Gamma_{b} ] = - \tfrac{i}{2} \Gamma_{a} \Gamma_{b}\,,a < b$,\\\\
b) four generators of the conformal boosts: $ \mathrm{K}_{a} = \tfrac{1}{2} \Gamma_{a}$,\\\\
c) four generators of the local translations: $ \mathrm{P}_{a} = -\tfrac{i}{2} \Gamma_{a} \Gamma_{5}$,\\\\
d) one generator for special conformal transformations: $\mathrm{D} = -\tfrac{1}{2} \Gamma_{5}$ and\\\\
e) one U(1) generator: $\one$. \\\\ 
The $\Gamma$-matrices are determined as tensor products of the Pauli matrices, specifically:
$$ \Gamma_{1} = \sigma_{1} \otimes \sigma_{1}, \ \ \ \Gamma_{2} = \sigma_{1} \otimes \sigma_{2}, \ \ \ \Gamma_{3} = \sigma_{1} \otimes \sigma_{3} $$
$$ \Gamma_{4} = \sigma_{2} \otimes \mathbf{1}, \ \ \  \Gamma_{5} = \sigma_{3} \otimes \mathbf{1}\,. $$
Therefore, the generators of the algebra are represented by the following 4$\times$4 matrices:
\begin{equation}
M_{ij} = - \frac{i}{2}\Gamma_{i} \Gamma_{j} = \frac{1}{2} \one \otimes \sigma_{k}\,,
\end{equation}
where $i,j,k = 1,2,3$ and:
\begin{equation}
M_{4k} = - \frac{i}{2}\Gamma_{4} \Gamma_{k} = - \frac{1}{2} \sigma_{3} \otimes \sigma_{k}\,.
\end{equation}
Straightforward calculations lead to the following commutation relations, which the operators satisfy:
\begin{eqnarray}
&&[ K_{a} , K_{b} ] = i M_{ab}, \ \ \ [P_{a}, P_{b} ] = i M_{ab} \nonumber \\
&&[ X_{a}, P_{b} ]  = i \delta_{ab}D, \ \ \ [X_{a}, D ] = i P_{a} \nonumber \\
&&[P_{a}, D ] =i K_{a} , \ \ \ [K_a,P_b]=i\delta_{ab}D , \ \ \ [K_a,D]=-iP_a \nonumber \\
&&[K_{a}, M_{bc} ] = i( \delta_{ac} K_{b} - \delta_{ab} K_{c} ) \nonumber \\
&&[P_{a}, M_{bc} ] = i( \delta_{ac} P_{b} - \delta_{ab} P_{c} ) \nonumber \\
&&[M_{ab}, M_{cd} ] = i( \delta_{ac} M_{bd} + \delta_{bd} M_{ac} - \delta_{bc}M_{ad} - \delta_{ad}M_{bc} ) \nonumber \\
&&[D, M_{ab} ] = 0\,.\label{algebra}
\end{eqnarray}

\subsection{Noncommutative gauge theory of gravity}
Having determined the commutation relations of the generators of the algebra, the noncommutative gauging procedure can be initiated. First, the covariant coordinate is defined as:
\begin{equation}
\hat{X}_m=X_m\otimes\one+A_m(X)\,.\label{covcoorddef}
\end{equation}
The property of covariance of the coordinate $\hat{X}_m$, is expressed as:
\begin{equation}
\delta\hat{X}_m=i[\epsilon,\hat{X}_m]\,,\label{covcoord}
\end{equation}
where $\epsilon(X)$ is the gauge transformation parameter. It is a function of the coordinates (N$\times$N matrices), $X_m$, but also is valued in the SO(6)$\times$U(1) algebra. Therefore, it can be decomposed on the sixteen generators of the algebra:
\begin{equation}
\epsilon=\epsilon_0(X)\otimes\one+\xi^a(X)\otimes K_a+\tilde{\epsilon}_0(X)\otimes D+\lambda_{ab}(X)\otimes\Sigma^{ab}+\tilde{\xi}^a(X)\otimes P_a\,.
\end{equation}
Taking into account that a gauge transformation acts trivially on the coordinate $X_m$, namely $\delta X_m=0$, the transformation property of the $A_m$ is obtained by the combination of the equations \eqref{covcoorddef} and \eqref{covcoord}. In accordance to the corresponding procedure in the commutative case, the $A_m$ transforms in such a way admitting the interpretation of the connection of the gauge theory. Similarly to the case of the gauge transformation parameter, $\epsilon$, the $A_m$, is a function of the coordinates $X_m$ of the fuzzy space $\mathrm{dS}_4$, but also takes values in the SO(6)$\times$U(1) algebra, which means that it can be expanded on its sixteen generators as follows:
\begin{equation}
A_m(X)=e_m^{~a}(X)\otimes P_a+\omega_{m}^{~ab}(X) \otimes \Sigma_{ab}(X) + b_{m}^{~a}(X) \otimes K_{a}(X) + \tilde{a}_{m}(X) \otimes D + a_{m}(X) \otimes  \one\,,
\end{equation}  
where it becomes manifest that one gauge field has been corresponded to each generator. The component gauge fields are functions of the coordinates of the space, $X_m$, therefore they have the form of N$\times$N matrices, where N is the dimension of the representation in which the  coordinates are accommodated. Therefore, instead of the ordinary product, between the gauge fields and their corresponding generators, the tensor product is used, since the factors are matrices of different dimensions, since the generators are represented by 4$\times$4 matrices. Therefore, it is concluded that each term in the expression of the gauge connection is a 4N$\times$4N matrix. 

After the decomposition of the gauge connection and the introduction of the gauge fields, the covariant coordinate is now written as:
\begin{equation}
\hat{X}_{m} = X_{m} \otimes \one + e_{m}^{~a}(X) \otimes P_{a} + \omega_{m}^{~ab}(X) \otimes \Sigma_{ab} + b_{m}^{~a} \otimes K_{a} + \tilde{a}_{m} \otimes D + a_{m} \otimes  \one\,.
\end{equation}
The next step is to calculate the field strength tensor for this SO(6)$\times$U(1) noncommutative gauge theory, which, for the fuzzy de Sitter space, is defined as:
\begin{equation}
\mathcal{R}_{mn} = [\hat{X}_{m}, \hat{X}_{n}]  - \frac{i\lambda^2}{\hbar}\hat{\Theta}_{mn}\,,\label{fieldstrengtt}
\end{equation}
where $\hat{\Theta}_{mn}=\Theta_{mn}\otimes\one+\mathcal{B}_{mn}$. The $\mathcal{B}_{mn}$ is a 2-form gauge field, which takes values in the SO(6)$\times$U(1) algebra. The $\mathcal{B}_{mn}$ field was introduced in order to make the field strength tensor covariant, since in its absence it does not transform covariantly\footnote{Details on this generic issue on such spaces are given in Appendix A of \cite{Manolakos:2019fle}.}. The $\mathcal{B}_{mn}$ field will contribute in the total action of the theory with a kinetic term of the following form:
\begin{equation}
\mathcal{S}_{\mathcal{B}}=\text{Tr}\,\text{tr}\, \hat{\mathcal{H}}_{mnp}\hat{\mathcal{H}}^{mnp}~.
\end{equation}
The $\hat{\mathcal{H}}_{mnp}$ field strength tensor transforms covariantly under a gauge transformation, therefore the above action is gauge invariant.

The field strength tensor of the gauge connection, \eqref{fieldstrengtt}, can be expanded in terms of the component curvature tensors, since it is valued in the algebra:
\begin{equation}
\mathcal{R}_{mn}(X) = R_{mn}^{~~~ab}(X) \otimes \Sigma_{ab} + \tilde{R}_{mn}^{~~a}(X) \otimes P_{a} + R_{mn}^{~~a}(X) \otimes K_{a} + \tilde{R}_{mn}(X) \otimes D + R_{mn}(X) \otimes \one\,.
\end{equation}
All necessary information for the determination of the transformations of the gauge fields and the expressions of the component curvature tensors is obtained. The explicit expressions and calculations lie in ref.\cite{Manolakos:2019fle}. 

\subsection{The constraints for the symmetry breaking and the action}

The gauge symmetry of the resulting theory, with which we would like to end up, is the one described by the Lorentz group, in the Euclidean signature, the SO(4). In this direction, one could consider directly a constrained theory in which the only component curvature tensors that would not be imposed to vanish would be the ones that corresponds to the Lorentz and the U(1) generators of the algebra, achieving a breaking of the initial SO(6)$\times$U(1) symmetry to the SO(4)$\times$U(1). However, counting the degrees of freedom, adoption of the above breaking would lead to an over-constrained theory. Therefore, it is more efficient to follow a different procedure and perform the symmetry breaking in a less straightforward way \cite{Manolakos:2019fle}. Accordingly, the first constraint is the torsionless condition:
\begin{equation}
\tilde{R}_{mn}^{~a}(P)=0\,, \label{ncconstraints}
\end{equation}
which is also encountered in the cases in which the Einstein and conformal gravity theories are described as (ordinary) gauge theories. Next, the gauge field $b_m^{~a}$ would admit an interpretation of a second vielbein of the theory, which would lead to a bimetric theory, which is not our case of interest. Thus, the relation $e_m^{~a}={b}_m^{~a}$ in the solution of the constraint should be considered. Also, this leads to the expression of the spin connection $\omega_m^{~ab}$ in terms of the rest of the independent fields, ${e}_m^{~a}, {a}_m, {\tilde{a}}_m$. In order to obtain the explicit expression of the spin connection in terms of the other fields, the following two identities are employed:
\begin{equation}
\delta^{abc}_{fgh}=\epsilon^{abcd}\epsilon_{fghd}\quad\quad\text{and}\quad\quad \frac{1}{3!}\delta^{abc}_{fgh}a^{fgh}=a^{[fgh]}\,.\label{ids}
\end{equation} 
Solving the constraint $\tilde{R}(P)=0$, one obtains:
\begin{equation}
\epsilon^{abcd}[e_m^{~b},\omega_n^{~cd}]-i\{\omega_m^{~ab},e_{nb}\}=-[D_m,e_m^{~a}]-i\{e_m^{~a},\tilde{a}_m\}\,,
\end{equation}
where $D_m=X_m+a_m$ being the covariant coordinate of an Abelian noncommutative gauge theory. The above equation leads to the following two:
\begin{equation}
\epsilon^{abcd}[e_m^{~b},\omega_n^{~cd}]=-[D_m,e_m^{~a}]\quad\quad \text{and}\quad\quad \{\omega_m^{~ab},e_{nb}\}=\{e_m^{~a},\tilde{a}_n\}\,.
\end{equation}
Taking into consideration the identities, \eqref{ids}, the above equations lead to the desired expression for the spin connection:
\begin{equation}
\omega_n^{~ac}=-\frac{3}{4}e^m_{~b}(-\epsilon^{abcd}[D_m,e_{nd}]+\delta^{[bc}\{e_n^{~a]},\tilde{a}_m\})\,.\label{omegaintermsofe}
\end{equation}
Next, according to ref.\cite{Green:1987mn}, the vanishing of the field strength tensor in a gauge theory could lead to the vanishing of the associated gauge field. However, the vanishing of the torsion component tensor, $\tilde{R}(P)=0$, does not imply $e_\mu^{~a}=0$, because such a choice would lead to degeneracy of the metric tensor of the space \cite{Witten:1988hc}. The field that can be gauge-fixed to zero is the $\tilde{a}_m$. This fixing, $\tilde{a}_m=0$, will modify the expression of the spin connection, \eqref{omegaintermsofe}, leading to a more simplified expression:
\begin{equation}
\omega_n^{~ac}=\frac{3}{4}e^m_{~b}\epsilon^{abcd}[D_m,e_{nd}]\,.\label{omegaintermsofeteliko}
\end{equation}     
It should be noted that the U(1) field strength tensor, $R_{mn}(\one)$, which is associated to the noncommutativity, is not considered to be vanishing. The symmetry breaking does not affect the U(1) part of the symmetry since the breaking takes place in the (high-energy) noncommutative regime. However, the corresponding field, $a_m$, decouples in the commutative limit of the broken theory. Therefore, in the commutative limit, the gauge symmetry would be just SO(4).    

Alternatively, another way to break the SO(6) gauge symmetry to the desired SO(4) is to induce a spontaneous symmetry breaking by including two scalar fields in the 6 representation of SO(6),  extrapolating the argument developed for the case of the conformal gravity to the noncommutative framework. It is expected that the spontaneous symmetry breaking induced by the scalars would lead to a constrained theory as the one that was obtained above by the imposition of the constraints \eqref{ncconstraints}. 

\subsection{The action and equations of motion}

Next, for the action of the theory, it is natural to consider one of Yang-Mills type\footnote{A Yang-Mills action $\text{tr}F^2$ defined on the fuzzy $\mathrm{dS}_4$ space is gauge invariant, for details see Appendix A of \cite{Manolakos:2019fle}.}:
\begin{equation}
\mathcal{S}=\text{Tr}\text{tr}\{ \mathcal{R}_{mn},\mathcal{R}_{rs}\}\epsilon^{mnrs}\,,
\end{equation}
where $\text{Tr}$ denotes the trace over the coordinates-N$\times$N matrices (it replaces the integration of the continuous case) and $\text{tr}$ denotes the trace over the generators of the algebra. After the symmetry breaking, that is including the constraints, the surviving terms of the action will be:
\begin{equation}
\mathcal{S}=2\text{Tr}(R_{mn}^{~~~ab}R_{rs}^{~~cd}\epsilon_{abcd}\epsilon^{mnrs}+4\tilde{R}_{mn}R_{rs}\epsilon^{mnrs}+\frac{1}{3}H_{mnp}^{~~~~ab}H^{mnpcd}\epsilon_{abcd}+\frac{4}{3}\tilde{H}_{mnp}H^{mnp})\,.
\end{equation}

Replacing with the explicit expressions of the component tensors and writing the $\omega$ gauge field in terms of the surviving gauge fields, \eqref{omegaintermsofeteliko}, then variation with respect to the independent gauge fields would lead to the equations of motion. 
\section{Conclusions}
In this review the construction of noncommutative 3-d and 4-d gravity models as gauge theories were revisited. Although for both cases the main procedure is similar, since they are based on the corresponding works in the continuous regime, let us summarize and conclude separately for the two cases.

In the 3-d case, the noncommutative background space we employed is the $\mathbb{R}_\lambda^3$ which is a foliation of the 3-d Euclidean space by multiple fuzzy spheres. This onion-like construction admits an SO(4) symmetry which is the one we chose as gauge group. The anticommutators of the generators of the gauge group would not close, that is why we promoted the symmetry to the U(2)$\times$U(2) and fixed its representation. Then, following the standard procedure, we defined the covariant coordinate and calculated the transformations of the fields and the expressions of the component curvature tensors. Naturally, the action we determined was of Chern-Simons type and the equations of motion were obtained after its variation. The results obtained in the above construction reduce to the ones of the commutative case. 

In the 4-d case, the noncommutative background space we employed is the fuzzy version of the 4-d de Sitter space. It is worth-mentioning that it consists a 4-d covariant noncommutative space, respecting Lorentz invariance, which is of major importance in our case. Next, we determined the gauge group, SO(5), which was enlarged for the same reasons as in the previous case, to the SO(6)$\times$U(1). Then, we went on following the standard procedure for the calculation of the transformations of the fields and the expressions of the component curvature tensors. Since we desired to result with a theory respecting the Lorentz symmetry, we imposed certain constraints in order to break the initial symmetry. After the symmetry breaking, the action takes its final form and its variation will lead to the equations of motion. The latter will be part of our future work. It should be noted that, before the symmetry breaking, the results of the above construction reduce to the ones of the conformal gravity in the commutative limit. Finally, it should be also stressed that the above is a matrix model giving insight into the gravitational interaction in the high-energy regime and also giving promises for improved UV properties as compared to ordinary gravity. Clearly, the latter, as well the inclusion of matter fields is going to be a subject of further study.

Another possible direction of further investigation is to consider our construction embedded in the Lie superalgebra of SU(2,2/1) which is isomorphic to the algebra of the superconformal spacetime symmetry group \cite{Yates:2017zgl}. Gauging of the latter leads to $\mathcal{N}=1$ conformal supergravity \cite{Fradkin:1985am,Kaku:1977pa,Kaku:1978nz,Kaku:1977rk,Townsend:1979ki}. In the present noncommutative context it seems more natural pursuing the gauging of the supergroup of SU(2,2/1) algebra. This possibility appeared to be fruitful in relating the Connes-Lott model \cite{connes-lott} to those based on gauging the SU(2/1) superalgebra \cite{Neeman:1979wp}-\cite{Hussain:1991tw} (see also \cite{Batakis:1993cw}) and could be useful in our construction, too.  
\\\\
\noindent {\bf Acknowledgements}:\\

\noindent We would like to thank Ali Chamseddine, Paolo Aschieri, Thanassis 
Chatzistavrakidis, Evgeny Ivanov, Larisa Jonke, Danijel Jurman, Alexander Kehagias, Dieter L\"ust, Denjoe O'Connor, Emmanuel Saridakis, Harold Steinacker, Kelly Stelle, Patrizia Vitale and Christof Wetterich for useful discussions.
The work of two of us (GM and GZ) was partially supported by the
COST Action MP1405, while both would like to thank ESI - Vienna for the 
hospitality during their participation in the Workshop ``Matrix Models for 
Noncommutative Geometry and String Theory'', Jul 09 - 13, 2018.
One of us (GZ) has been supported within the Excellence Initiative funded 
by the German and States Governments, at the Institute for Theoretical 
Physics, Heidelberg University and from the Excellent Grant Enigmass
of LAPTh. GZ would like to thank the ITP - Heidelberg, LAPTh - Annecy and 
MPI - Munich for their hospitality.


\begin{thebibliography}{99}
\bibitem{Utiyama:1956sy}
  R.~Utiyama,
  Phys.\ Rev.\  {\bf 101} (1956) 1597.
   doi:10.1103/PhysRev.101.1597

\bibitem{Kibble:1961ba}
  T.~W.~B.~Kibble,
  J.\ Math.\ Phys.\  {\bf 2} (1961) 212.
  doi:10.1063/1.1703702

\bibitem{Stelle:1979aj}
  K.~S.~Stelle and P.~C.~West,
  Phys.\ Rev.\ D {\bf 21} (1980) 1466.
  doi:10.1103/PhysRevD.21.1466

\bibitem{MacDowell:1977jt}
  S.~W.~MacDowell and F.~Mansouri,
  Phys.\ Rev.\ Lett.\  {\bf 38} (1977) 739
   Erratum: [Phys.\ Rev.\ Lett.\  {\bf 38} (1977) 1376].
 doi:10.1103/PhysRevLett.38.1376, 10.1103/PhysRevLett.38.739

\bibitem{Ivanov:1980tw}
  E.~A.~Ivanov and J.~Niederle, Conference: C80-06-23.3, p.545-551, 1980; 
  E.~A.~Ivanov and J.~Niederle,
  Phys.\ Rev.\ D {\bf 25} (1982) 976.
  doi:10.1103/PhysRevD.25.976; 
  E.~A.~Ivanov and J.~Niederle,
  Phys.\ Rev.\ D {\bf 25} (1982) 988.
  doi:10.1103/PhysRevD.25.988

\bibitem{Kibble:1985sn}
  T.~W.~B.~Kibble and K.~S.~Stelle,
  In Ezawa, H. ( Ed.), Kamefuchi, S. ( Ed.): Progress In Quantum Field Theory, 57-81.

\bibitem{Kaku:1977pa}
  M.~Kaku, P.~K.~Townsend and P.~van Nieuwenhuizen,
  ``Gauge Theory of the Conformal and Superconformal Group,''
  Phys.\ Lett.\  {\bf 69B} (1977) 304.
  doi:10.1016/0370-2693(77)90552-4

\bibitem{Fradkin:1985am}
  E.~S.~Fradkin and A.~A.~Tseytlin,
  ``Conformal Supergravity,''
  Phys.\ Rept.\  {\bf 119} (1985) 233.
  doi:10.1016/0370-1573(85)90138-3

\bibitem{vanproeyen}
  D.~Z.~Freedman and A.~Van~Proeyen
  ``Supergravity,'' Cambridge University Press, 2012

\bibitem{cham-thesis}
A. H. Chamseddine, ``Supersymmetry and higher spin fields'', PhD Thesis, (1976)

\bibitem{Chamseddine:1976bf}
  A.~H.~Chamseddine and P.~C.~West,
  Nucl.\ Phys.\ B {\bf 129} (1977) 39.
  doi:10.1016/0550-3213(77)90018-9

\bibitem{Witten:1988hc}
  E.~Witten,
  ``(2+1)-Dimensional Gravity as an Exactly Soluble System,''
  Nucl.\ Phys.\ B {\bf 311} (1988) 46.
  

\bibitem{connes}
Connes A., Academic Press, Inc., San
Diego, CA, 1994.


\bibitem{madorej}
Madore J., London Mathematical
Society Lecture Note Series, Vol. 257, Cambridge University Press,
Cambridge, 1999.


\bibitem{Madore:1991bw}
  J.~Madore,
  Class.\ Quant.\ Grav.\  {\bf 9} (1992) 69.
  doi:10.1088/0264-9381/9/1/008



\bibitem{buric-grammatikopoulos-madore-zoupanos}
Buric M., Grammatikopoulos T., Madore J., Zoupanos G., JHEP 0604 (2006) 054;
Buric M., Madore J., Zoupanos G., SIGMA 3:125,2007, arXiv:0712.4024
[hep-th].


\bibitem{filk}
T. Filk, Phys.
Lett. B 376 (1996) 53; J. C. V\'{a}rilly and J. M.
Gracia-Bond\'{i}a, Int. J. Mod. Phys. A 14 (1999) 1305
[hep-th/9804001]; M. Chaichian, A. Demichev and P. Presnajder,
Nucl. Phys. B 567 (2000)
360, hepth/ 9812180; S. Minwalla, M. Van Raamsdonk and N. Seiberg,
JHEP 0002 (2000) 020, hep-th/9912072.


\bibitem{grosse-wulkenhaar}
H. Grosse and R. Wulkenhaar, Lett. Math. Phys. 71
(2005) 13, hep-th/0403232.


\bibitem{grosse-steinacker}
H. Grosse and H. Steinacker, Adv. Theor. Math.
Phys. 12 (2008) 605, hep-th/0607235; H. Grosse and H. Steinacker,
Nucl. Phys. B 707 (2005)
145, hep-th/0407089.


\bibitem{connes-lott}
Connes A., Lott J., Nuclear Phys. B Proc. Suppl. 18 (1991), 29-47; Chamseddine A.H.,
Connes A., Comm. Math. Phys. 186
(1997), 731-750, hep-th/9606001; Chamseddine A.H., Connes A.,
Phys. Rev. Lett. 99 (2007),
191601, arXiv:0706.3690.


\bibitem{martin-bondia}
Mart\'{i}n C.P., Gracia-Bond{\'i}a M.J., V{\'a}rilly J.C., Phys. Rep. 294 (1998), 363-406, hep-th/9605001.


\bibitem{dubois-madore-kerner}
Dubois-Violette M., Madore J., Kerner R., Phys. Lett. B217 (1989), 485-488;
Dubois-Violette M., Madore J., Kerner R., Classical Quantum Gravity 6 (1989),
1709-1724; Dubois-Violette M., Kerner R., Madore J.,
J. Math. Phys. 31 (1990), 323-330.


\bibitem{madorejz}
Madore J., Phys. Lett. B 305 (1993),
84-89; Madore J., (Sobotka Castle, 1992), Fund. Theories Phys., Vol.
52, Kluwer Acad. Publ., Dordrecht, 1993, 285-298. hep-ph/9209226.


\bibitem{connes-douglas-schwarz}
Connes A., Douglas M.R., Schwarz A., JHEP (1998), no.2, 003,
hep-th/9711162.


\bibitem{seiberg-witten}
Seiberg N., Witten E., JHEP (1999), no.9, 032, hep-th/9908142.


\bibitem{ishibasi-kawai}
N.Ishibashi, H.Kawai, Y.Kitazawa and A.Tsuchiya, Nucl. Phys. B498 (1997) 467, arXiv:hep-th/9612115.


\bibitem{jurco}
Jur\v{c}o B., Schraml S., Schupp P., Wess J., Eur. Phys. J. C 17 (2000), 521-526,
hep-th/0006246; Jur\v{c}o B., Schupp P., Wess J.,
Nuclear Phys. B 604 (2001), 148-180, hep-th/0102129; Jur{\v c}o B.,
Moller L., Schraml S., Schupp S., Wess J., Eur. Phys. J.
C 21 (2001), 383-388, hep-th/0104153; Barnich G., Brandt F.,
Grigoriev M., JHEP (2002), no.8, 023,
hep-th/0206003.


\bibitem{chaichian}
Chaichian M., Pre{\v s}najder P., Sheikh-Jabbari M.M., Tureanu A.,
Eur. Phys. J. C
29 (2003), 413-432, hep-th/0107055.


\bibitem{camlet}
Calmet X., Jur{\v c}o B., Schupp P., Wess J., Wohlgenannt M., Eur. Phys. J. C 23
(2002), 363-376, hep-ph/0111115; Aschieri P., Jur{\v c}o B., Schupp
P., Wess J., Nuclear Phys. B 651 (2003), 45-70, hep-th/0205214; Behr W.,
Deshpande N.G., Duplancic G., Schupp P., Trampetic J., Wess J.,
Eur.Phys.J.C29: 441-446, 2003.


\bibitem{aschieri-madore-manousselis-zoupanos}
Aschieri P., Madore J., Manousselis P., Zoupanos G., JHEP (2004), no. 4, 034,
hep-th/0310072; Aschieri P., Madore J., Manousselis P., Zoupanos G.,
Fortschr. Phys. 52
(2004), 718-723, hep-th/0401200; Aschieri P., Madore J., Manousselis
P., Zoupanos G., Conference: C04-08-20.1 (2005) 135-146,
hep-th/0503039.


\bibitem{aschieri-grammatikopoulos}
Aschieri P., Grammatikopoulos T., Steinacker H., Zoupanos G.,
JHEP (2006), no. 9, 026,
hep-th/0606021; Aschieri P., Steinacker H., Madore J., Manousselis
P., Zoupanos G., arXiv:0704.2880.


\bibitem{steinacker-zoupanos}
Steinacker H., Zoupanos G., JHEP (2007), no. 9, 017,
arXiv:0706.0398.


\bibitem{chatzistavrakidis-steinacker-zoupanos}
A. Chatzistavrakidis, H. Steinacker and G. Zoupanos, Fortsch.Phys. 58 (2010)
537-552, arXiv:0909.5559 [hep-th].


\bibitem{fuzzy}
A.~Chatzistavrakidis, H.~Steinacker and G.~Zoupanos,
    JHEP 1005 (2010) 100,
  arXiv:hep-th/1002.2606
A.~Chatzistavrakidis and G.~Zoupanos,
     SIGMA 6 (2010) 063, arXiv:hep-th/1008.2049.


\bibitem{Gavriil:2015lka}
  D.~Gavriil, G.~Manolakos, G.~Orfanidis and G.~Zoupanos,
  Fortsch.\ Phys.\  {\bf 63} (2015) 442
  doi:10.1002/prop.201500022
  [arXiv:1504.07276 [hep-th]]; 
  G.~Manolakos and G.~Zoupanos,
  Phys.\ Part.\ Nucl.\ Lett.\  {\bf 14} (2017) no.2,  322.
  doi:10.1134/S1547477117020194; 
  G.~Manolakos and G.~Zoupanos,
  Springer Proc.\ Math.\ Stat.\  {\bf 191} (2016) 203
  doi:10.1007/978-981-10-2636-2-13
  [arXiv:1602.03673 [hep-th]].
  
\bibitem{Madore:2000en}
  J.~Madore, S.~Schraml, P.~Schupp and J.~Wess,
  Eur.\ Phys.\ J.\ C {\bf 16} (2000) 161
  doi:10.1007/s100520050012
  [hep-th/0001203].

\bibitem{Chamseddine:2000si}
  A.~H.~Chamseddine,
  ``Deforming Einstein's gravity,''
  Phys.\ Lett.\ B {\bf 504} (2001) 33
  doi:10.1016/S0370-2693(01)00272-6
  [hep-th/0009153].
  
\bibitem{Chamseddine:2003we}
A.~H.~Chamseddine,
Phys.\ Rev.\ D {\bf 69} (2004) 024015
doi:10.1103/PhysRevD.69.024015
[hep-th/0309166].

\bibitem{Aschieri1}
P.~Aschieri and L.~Castellani,
JHEP {\bf 0906} (2009) 086
doi:10.1088/1126-6708/2009/06/086
[arXiv:0902.3817 [hep-th]].

\bibitem{Aschieri2}
P.~Aschieri and L.~Castellani,
JHEP {\bf 0906} (2009) 087
doi:10.1088/1126-6708/2009/06/087
[arXiv:0902.3823 [hep-th]].

\bibitem{Ciric:2016isg}
M.~Dimitrijevi\'c \'Ciri\'c, B.~Nikoli\'c and V.~Radovanovi\'c,
Phys.\ Rev.\ D {\bf 96} (2017) no.6,  064029
doi:10.1103/PhysRevD.96.064029
[arXiv:1612.00768 [hep-th]].
  
\bibitem{Cacciatori:2002gq}
S.~Cacciatori, D.~Klemm, L.~Martucci and D.~Zanon,
Phys.\ Lett.\ B {\bf 536} (2002) 101
doi:10.1016/S0370-2693(02)01823-3
[hep-th/0201103].

\bibitem{Cacciatori:2002ib}
S.~Cacciatori, A.~H.~Chamseddine, D.~Klemm, L.~Martucci, W.~A.~Sabra and D.~Zanon,
Class.\ Quant.\ Grav.\  {\bf 19} (2002) 4029 
doi:10.1088/0264-9381/19/15/310
[hep-th/0203038].

\bibitem{Aschieri3}
P.~Aschieri and L.~Castellani,
JHEP {\bf 1411} (2014) 103
doi:10.1007/JHEP11(2014)103 
[arXiv:1406.4896 [hep-th]].

\bibitem{Banados:2001xw}
  M.~Banados, O.~Chandia, N.~E.~Grandi, F.~A.~Schaposnik and G.~A.~Silva,
  Phys.\ Rev.\ D {\bf 64} (2001) 084012
  doi:10.1103/PhysRevD.64.084012
  [hep-th/0104264].
  
\bibitem{Seiberg:1999vs}
N.~Seiberg and E.~Witten,
JHEP {\bf 9909} (1999) 032
doi:10.1088/1126-6708/1999/09/032
[hep-th/9908142].

\bibitem{Banks:1996vh}
T.~Banks, W.~Fischler, S.~H.~Shenker and L.~Susskind,
Phys.\ Rev.\ D {\bf 55} (1997) 5112
doi:10.1103/PhysRevD.55.5112
[hep-th/9610043].

\bibitem{Ishibashi:1996xs}
N.~Ishibashi, H.~Kawai, Y.~Kitazawa and A.~Tsuchiya,
Nucl.\ Phys.\ B {\bf 498} (1997) 467
doi:10.1016/S0550-3213(97)00290-3
[hep-th/9612115].
  
\bibitem{Aoki:1998vn}
H.~Aoki, S.~Iso, H.~Kawai, Y.~Kitazawa and T.~Tada,
Prog.\ Theor.\ Phys.\  {\bf 99} (1998) 713
doi:10.1143/PTP.99.713
[hep-th/9802085].

\bibitem{Hanada:2005vr}
M.~Hanada, H.~Kawai and Y.~Kimura,
Prog.\ Theor.\ Phys.\  {\bf 114} (2006) 1295
doi:10.1143/PTP.114.1295
 [hep-th/0508211].

\bibitem{Furuta:2006kk}
K.~Furuta, M.~Hanada, H.~Kawai and Y.~Kimura,
Nucl.\ Phys.\ B {\bf 767} (2007) 82
doi:10.1016/j.nuclphysb.2007.01.003
 [hep-th/0611093].

\bibitem{Yang:2006dk}
H.~S.~Yang,
Int.\ J.\ Mod.\ Phys.\ A {\bf 24} (2009) 4473
doi:10.1142/S0217751X0904587X
[hep-th/0611174].

\bibitem{Steinacker:2010rh}
H.~Steinacker,
Class.\ Quant.\ Grav.\  {\bf 27} (2010) 133001
doi:10.1088/0264-9381/27/13/133001
[arXiv:1003.4134 [hep-th]].

\bibitem{Kim:2011cr}
S.~W.~Kim, J.~Nishimura and A.~Tsuchiya,
Phys.\ Rev.\ Lett.\  {\bf 108} (2012) 011601
doi:10.1103/PhysRevLett.108.011601
[arXiv:1108.1540 [hep-th]].

\bibitem{Nishimura:2012xs}
J.~Nishimura,
PTEP {\bf 2012} (2012) 01A101
doi:10.1093/ptep/pts004
[arXiv:1205.6870 [hep-lat]].

\bibitem{Nair:2001kr}
V.~P.~Nair,
Nucl.\ Phys.\ B {\bf 651} (2003) 313
doi:10.1016/S0550-3213(02)01061-1
[hep-th/0112114].

\bibitem{Abe:2002in}
Y.~Abe and V.~P.~Nair,
Phys.\ Rev.\ D {\bf 68} (2003) 025002
doi:10.1103/PhysRevD.68.025002
[hep-th/0212270].

\bibitem{Valtancoli:2003ve}
P.~Valtancoli,
Int.\ J.\ Mod.\ Phys.\ A {\bf 19} (2004) 361
doi:10.1142/S0217751X04017598
[hep-th/0306065].

\bibitem{Nair:2006qg}
V.~P.~Nair,
Nucl.\ Phys.\ B {\bf 750} (2006) 321
doi:10.1016/j.nuclphysb.2006.06.009
[hep-th/0605008].

\bibitem{Buric:2006di}
M.~Buri\'c, T.~Grammatikopoulos, J.~Madore and G.~Zoupanos,
JHEP {\bf 0604} (2006) 054
doi:10.1088/1126-6708/2006/04/054
[hep-th/0603044].

\bibitem{Buric:2007zx}
M.~Buri\'c, J.~Madore and G.~Zoupanos,
SIGMA {\bf 3} (2007) 125
doi:10.3842/SIGMA.2007.125
[arXiv:0712.4024 [hep-th]].

\bibitem{Buric:2007hb}
M.~Buri\'c, J.~Madore and G.~Zoupanos,
Eur.\ Phys.\ J.\ C {\bf 55} (2008) 489
doi:10.1140/epjc/s10052-008-0602-x
[arXiv:0709.3159 [hep-th]].

\bibitem{Aschieri:2003vyAschieri:2004vhAschieri:2005wm}
  P.~Aschieri, J.~Madore, P.~Manousselis and G.~Zoupanos,
  JHEP {\bf 0404} (2004) 034
  doi:10.1088/1126-6708/2004/04/034
  [hep-th/0310072];
  ibid,  
  Fortsch.\ Phys.\  {\bf 52} (2004) 718
  doi:10.1002/prop.200410168
  [hep-th/0401200];
  ibid,
  hep-th/0503039.

\bibitem{Snyder:1946qz}
  H.~S.~Snyder,
  Phys.\ Rev.\  {\bf 71} (1947) 38.
  doi:10.1103/PhysRev.71.38

\bibitem{Yang:1947ud}
  C.~N.~Yang,
  Phys.\ Rev.\  {\bf 72} (1947) 874.
  doi:10.1103/PhysRev.72.874
  
\bibitem{Heckman:2014xha}
  J.~Heckman and H.~Verlinde,
  ``Covariant non-commutative spacetime,''
  Nucl.\ Phys.\ B {\bf 894} (2015) 58
  [arXiv:1401.1810 [hep-th]].
  
\bibitem{Buric:2015wta}
M.~Buri\'{c} and J.~Madore,
Eur.\ Phys.\ J.\ C {\bf 75} (2015) no.10,  502
doi:10.1140/epjc/s10052-015-3729-6
 [arXiv:1508.06058 [hep-th]].


\bibitem{Sperling:2017dts}
M.~Sperling and H.~C.~Steinacker,
J.\ Phys.\ A {\bf 50} (2017) no.37,  375202
doi:10.1088/1751-8121/aa8295
[arXiv:1704.02863 [hep-th]].

\bibitem{Buric:2017yes}
M.~Buri\'{c}, D.~Latas and L.~Nenadovi\'{c},
arXiv:1709.05158 [hep-th].
  
\bibitem{Steinacker:2016vgf}
  H.~C.~Steinacker,
  JHEP {\bf 1612} (2016) 156
  doi:10.1007/JHEP12(2016)156
  [arXiv:1606.00769 [hep-th]].

\bibitem{Chatzistavrakidis:2018vfi}
  A.~Chatzistavrakidis, L.~Jonke, D.~Jurman, G.~Manolakos, P.~Manousselis and G.~Zoupanos,
  Fortsch.\ Phys.\  {\bf 66} (2018) no.8-9,  1800047
  doi:10.1002/prop.201800047
  [arXiv:1802.07550 [hep-th]].

\bibitem{Manolakos:2018isw}
  D.~Jurman, G.~Manolakos, P.~Manousselis and G.~Zoupanos,
  PoS CORFU {\bf 2017} (2018) 162
  doi:10.22323/1.318.0162
  [arXiv:1809.03879 [gr-qc]].
 

\bibitem{Manolakos:2018hvn}
  G.~Manolakos and G.~Zoupanos,
  ``Non-commutativity in Unified Theories and Gravity,''
  Springer Proc.\ Math.\ Stat.\  {\bf 263} (2017) 177
  doi:10.1007/978-981-13-2715-5-10
  [arXiv:1809.02954 [hep-th]].


\bibitem{Hammou:2001cc}
A.~B.~Hammou, M.~Lagraa and M.~M.~Sheikh-Jabbari,
Phys.\ Rev.\ D {\bf 66} (2002) 025025
doi:10.1103/PhysRevD.66.025025
[hep-th/0110291]. 


\bibitem{Vitale:2014hca}
  P.~Vitale,
  Fortsch.\ Phys.\  {\bf 62} (2014) 825
  doi:10.1002/prop.201400037
  [arXiv:1406.1372 [hep-th]].
  

\bibitem{Kovacik:2013yca}
S.~Kov\'{a}\v{c}ik and P.~Pre\v{s}najder,
J.\ Math.\ Phys.\  {\bf 54} (2013) 102103
doi:10.1063/1.4826355
[arXiv:1309.4592 [math-ph]].


\bibitem{Jurman:2013ota}
D.~Jurman and H.~Steinacker,
JHEP {\bf 1401} (2014) 100
doi:10.1007/JHEP01(2014)100
[arXiv:1309.1598 [hep-th]].  

\bibitem{Manolakos:2019fle}
  G.~Manolakos, P.~Manousselis and G.~Zoupanos,
  arXiv:1902.10922 [hep-th].
  
\bibitem{Chamseddine:2002fd}
  A.~H.~Chamseddine,
  J.\ Math.\ Phys.\  {\bf 44} (2003) 2534
  doi:10.1063/1.1572199
  [hep-th/0202137].
  
\bibitem{Li:1973mq}
  L.~F.~Li,
  ``Group Theory of the Spontaneously Broken Gauge Symmetries,''
  Phys.\ Rev.\ D {\bf 9} (1974) 1723.
  doi:10.1103/PhysRevD.9.1723
  

\bibitem{Chamseddine:2002fd}
  A.~H.~Chamseddine,
  ``Invariant actions for noncommutative gravity,''
  J.\ Math.\ Phys.\  {\bf 44} (2003) 2534
  doi:10.1063/1.1572199
  [hep-th/0202137].
  
  
 \bibitem{DeBellis:2010sy}
 J.~DeBellis, C.~Saemann and R.~J.~Szabo,
 ``Quantized Nambu-Poisson Manifolds in a 3-Lie Algebra Reduced Model,''
 JHEP {\bf 1104} (2011) 075
 doi:10.1007/JHEP04(2011)075
 [arXiv:1012.2236 [hep-th]].  
 
\bibitem{Singh:2018qzk}
  A.~Singh and S.~M.~Carroll,
  ``Modeling Position and Momentum in Finite-Dimensional Hilbert Spaces via Generalized Clifford Algebra,''
  arXiv:1806.10134 [quant-ph].
  
  \bibitem{Barut}
  A.~Barut, `` From Heisenberg algebra to Conformal Dynamical Group ' '
  in A.~Barut, H.~D.~Doener (Eds) ``Conformal Groups and related Symmetries.Physical Results and Mathematical Backgrounf ''
  Lecture Notes in Physics, Springer-Verlag 1985  
  
\bibitem{AlvarezGaume:2006bn}
  L.~Alvarez-Gaume, F.~Meyer and M.~A.~Vazquez-Mozo,
  ``Comments on noncommutative gravity,''
  Nucl.\ Phys.\ B {\bf 753} (2006) 92
  doi:10.1016/j.nuclphysb.2006.07.009
  [hep-th/0605113].
 

\bibitem{Kimura:2002nq}
  Y.~Kimura,
  ``Noncommutative gauge theory on fuzzy four sphere and matrix model,''
  Nucl.\ Phys.\ B {\bf 637} (2002) 177
  doi:10.1016/S0550-3213(02)00469-8
  [hep-th/0204256].
  
\bibitem{Green:1987mn}
  M.~B.~Green, J.~H.~Schwarz and E.~Witten,
  ``Superstring Theory. Vol. 2: Loop Amplitudes, Anomalies And Phenomenology,''
  Cambridge, Uk: Univ. Pr. ( 1987) 596 P. ( Cambridge Monographs On Mathematical Physics)
  
\bibitem{Yates:2017zgl}
  L.~A.~Yates and P.~D.~Jarvis,
  J.\ Phys.\ A {\bf 51} (2018) no.14,  145203
  doi:10.1088/1751-8121/aab215
  [arXiv:1710.10533 [hep-th]].
  
  
\bibitem{Kaku:1978nz}
  M.~Kaku, P.~K.~Townsend and P.~van Nieuwenhuizen,
  Phys.\ Rev.\ D {\bf 17} (1978) 3179.
  doi:10.1103/PhysRevD.17.3179
  
\bibitem{Kaku:1977rk}
  M.~Kaku, P.~K.~Townsend and P.~van Nieuwenhuizen,
  Phys.\ Rev.\ Lett.\  {\bf 39} (1977) 1109.
  doi:10.1103/PhysRevLett.39.1109
  
\bibitem{Townsend:1979ki}
  P.~K.~Townsend and P.~van Nieuwenhuizen,
  Phys.\ Rev.\ D {\bf 19} (1979) 3166.
  doi:10.1103/PhysRevD.19.3166
  
\bibitem{Neeman:1979wp}
  Y.~Ne'eman,
  Phys.\ Lett.\ B {\bf 81} (1979) 190
   [Phys.\ Lett.\  {\bf 81B} (1979) 190].
  doi:10.1016/0370-2693(79)90521-5
  
\bibitem{Coquereaux:1990ev}
  R.~Coquereaux, G.~Esposito-Farese and G.~Vaillant,
  Nucl.\ Phys.\ B {\bf 353} (1991) 689.
  doi:10.1016/0550-3213(91)90323-P
  
\bibitem{Dondi:1979pb}
  P.~H.~Dondi and P.~D.~Jarvis,
  Z.\ Phys.\ C {\bf 4} (1980) 201.
  doi:10.1007/BF01421797
  
\bibitem{Hussain:1990nk}
  F.~Hussain and G.~Thompson,
  Phys.\ Lett.\ B {\bf 260} (1991) 359.
  doi:10.1016/0370-2693(91)91625-6
  
\bibitem{Hussain:1991tw}
  F.~Hussain and G.~Thompson,
  Phys.\ Lett.\ B {\bf 265} (1991) 307.
  doi:10.1016/0370-2693(91)90058-X
  
\bibitem{Batakis:1993cw}
  N.~A.~Batakis, A.~A.~Kehagias and G.~Zoupanos,
  Phys.\ Lett.\ B {\bf 315} (1993) 319
   Erratum: [Phys.\ Lett.\ B {\bf 318} (1993) 650].
  doi:10.1016/0370-2693(93)90470-3, 10.1016/0370-2693(93)91619-X
\end{thebibliography}
\end{document}